\documentclass{article} 
\usepackage{iclr2027_conference,times}

\usepackage{amsmath,amsfonts,bm}

\def\eqref#1{equation~\ref{#1}}

\def\1{\bm{1}}

\DeclareMathAlphabet{\mathsfit}{\encodingdefault}{\sfdefault}{m}{sl}
\SetMathAlphabet{\mathsfit}{bold}{\encodingdefault}{\sfdefault}{bx}{n}

\usepackage{xspace} 
\def\sysname{SoundWeaver\xspace}
\usepackage{graphicx}
\usepackage{hyperref}
\usepackage{url}
\usepackage{caption}     
\usepackage{adjustbox}   
\usepackage{tcolorbox}
\tcbuselibrary{skins,breakable}
\usepackage{multirow}    
\usepackage{booktabs}    
\usepackage{array}       
\usepackage{hyperref}
\usepackage{url}
\usepackage{graphics, enumitem}

\usepackage{booktabs}
\usepackage{multirow}
\usepackage[table]{xcolor}
\usepackage{adjustbox}

\usepackage{amsthm}
\newtheorem{proposition}{Proposition}

\definecolor{oursbg}{RGB}{247, 218, 218}

\newenvironment{denseitemize}{
\begin{itemize}[topsep=0pt, partopsep=0pt, leftmargin=1.5em]
  \setlength{\itemsep}{2pt}
  \setlength{\parskip}{0pt}
  \setlength{\parsep}{0pt}
}{\end{itemize}}

\usepackage{xcolor}

\hypersetup{
    colorlinks=true,
    citecolor=violet,
    linkcolor=violet,
    urlcolor=blue
}

\tcolorboxenvironment{proposition}{
    colback=black!2,
    colframe=black!35,
    boxrule=0.5pt,
    arc=1mm,
    left=1.5mm,
    right=1.5mm,
    top=1mm,
    bottom=1mm,
    before skip=6pt,
    after skip=6pt
}
\title{SoundWeaver: Compositional Warm-Starting for Text-to-Audio Diffusion Serving}

\author{%
Ayush Barik\textsuperscript{1},
Sofia Stoica\textsuperscript{1},
Nikhil Sarda\textsuperscript{2},
Arnav Kethana\textsuperscript{1}\\%
\textbf{Abhinav Khanduja\textsuperscript{1},
Muchen Xu\textsuperscript{1},
Fan Lai\textsuperscript{1}}\\[3pt]%
\normalfont
\textsuperscript{1}University of Illinois Urbana-Champaign
\qquad
\textsuperscript{2}Assured Intelligence\\%
\texttt{\{barik2,sstoica2,kethana2,AK89,muchenx2,fanlai\}@illinois.edu}\\%
\texttt{nikhilsarda.iitkgp@gmail.com}%
}
\iclrfinalcopy 
\begin{document}

\maketitle

\begin{abstract}
Text-to-audio (T2A) diffusion models generate high-quality audio but require
tens of neural function evaluations (NFEs), resulting in substantial inference
latency. Existing acceleration methods primarily optimize each generation
trajectory independently, even when reusable acoustic structure exists in prior
examples. We introduce \sysname, the first training-free framework for compositional cross-example acoustic reuse. Rather than generating from
pure noise, \sysname constructs a prompt-conditioned acoustic prior from
segments distributed across cached audios and warm-starts diffusion from an
intermediate noise level. We introduce an \emph{Acoustic Composition Graph}
(ACG), which uses neural-codec representations and segment-level identification to identify
candidate cross-cache transitions and structured path search to jointly
optimize prompt relevance and acoustic continuity. We further establish a
warm-start error bound showing how prior quality and diffusion SNR jointly
govern the admissible skip depth, motivating an adaptive contextual-bandit
\emph{Skip Gater} that selects the warm-start level for each request.
Across T2A backbones, \sysname achieves $1.48$--$3.74\times$
generation speedup with only a 1K-clip cache while preserving or improving
generation quality.
\end{abstract}

\section{INTRODUCTION}
Diffusion and flow-matching models have become the foundation of text-to-audio (T2A)
generation, such as for music composition and sound-effect
synthesis~\citep{audioldm2-2024taslp, liu2023audioldm}. These models
iteratively transform Gaussian noise into high-fidelity waveforms or
spectrograms over tens of sampling steps, conditioning on natural language
prompts. Despite their quality advantages, inference remains inherently
computationally intensive: generating a single clip can take multiple seconds
on modern GPUs~\citep{li2025meanaudiofastfaithfultexttoaudio}. At production
scale, where services process millions of requests
daily, this translates into high user-perceived latency
and substantial infrastructure cost.

Most existing acceleration methods reduce the cost of an individual
generation trajectory. Training-based approaches
distill multi-step diffusion into few-step generators, but require additional
training for each target model, 
which implies substantial computational cost and data resources~\citep{salimans2022progressive,pmlr-v202-song23a}.
Training-free approaches instead employ improved numerical
solvers~\citep{liu2022pseudo,NEURIPS2022_260a14ac}, adaptive timestep
schedules~\citep{chen2024adaptive}, or early-exit
strategies~\citep{pmlr-v235-moon24a}. Concurrently, systems work improves
execution through multi-GPU parallelism and hardware--software co-design~\citep{fang2024xditinferenceenginediffusion,
qiao2026flashdiff,agarwal2024nirvana,xia2025modm}.

We investigate a different source of efficiency: \emph{can acoustic structure from prior
examples be reused instead of repeatedly reconstructed from noise?}
Real-world audio contains substantial semantic recurrence---such as
rainfall, footsteps, engines, speech, or combinations of common acoustic
events. Our analysis shows that even a
modest 1K-audio cache frequently contains semantically related content
(Fig.~\ref{fig:motivation}).  At the same time, diffusion warm-starting provides
a mechanism for exploiting such structure: rather than initializing from pure
noise, a related audio prior can be noised to an intermediate timestep and
denoised only from that point onward, skipping part of the reverse trajectory~\citep{agarwal2024nirvana,xia2025modm}. We further show theoretically that a prior closer to the prompt-conditioned target
distribution can tolerate a less-noised warm start, allowing more of the
reverse process to be skipped (\S\ref{subsec:motivation}).

The central difficulty is that, for audio, the useful unit of reuse is rarely
an entire audio. Semantic relevance is temporally localized: a cached clip
may contain only part of the requested event sequence, mix relevant and
irrelevant intervals, or differ from the requested duration. Our analysis on real-world datasets shows that progressively partitioning audios into
shorter segments exposes substantially stronger prompt--audio matches than
whole-clip retrieval (\S\ref{fig:motivation}). This suggests moving beyond the single-reference
assumption used by prior warm-start systems for image generation~\citep{xia2025modm,agarwal2024nirvana}. Instead, a useful acoustic prior
may need to combine complementary structure distributed across different
times and different audios.

We introduce \sysname, a training-free framework for
\emph{compositional acoustic reuse}. Rather than retrieving one cached output,
\sysname constructs a prompt-conditioned acoustic prior from reusable segments
across audios. To make such composition feasible, we introduce an
\emph{Acoustic Composition Graph} (ACG). Its nodes represent cached audio
segments, while candidate cross-audio transitions are indexed using
coarse-to-fine neural-codec prefixes and scored for acoustic continuity.
Given a prompt and target duration, \sysname assigns segment-level semantic
scores and solves a dynamic program that jointly optimizes prompt relevance
and transition compatibility. A small set of high-scoring trajectories is
then assembled from the original waveforms and globally reranked against the
complete prompt, yielding the final acoustic prior.

A strong prior alone does not determine how aggressively it should be reused.
Preserving more of the prior allows more denoising steps to be skipped, but
also leaves less opportunity for the generative model to revise mismatched
content. \sysname introduces an adaptive \emph{Skip Gater} that
selects the warm-start timestep for each request. Guided by our warm-start
error analysis, the gater operates over log-SNR-spaced skip actions and uses a
contextual bandit conditioned on the prompt, composed prior, denoising budget,
and serving state. Together, ACG and the Skip Gater address the two sides of
warm-start acceleration: constructing more reusable acoustic structure and
determining how much of it can be preserved.

\textbf{Contributions.}
Overall, we make three key contributions:
\begin{denseitemize}
\item To our knowledge, \sysname is the first T2A inference framework to
accelerate generation through compositional cross-example acoustic
reuse. We formalize why a better acoustic prior permits a less-noised
warm start and hence greater denoising savings
(\S\ref{subsec:motivation}).

\item We introduce the \emph{Acoustic Composition Graph (ACG)}, which composes
prompt-relevant segments across caches while accounting for acoustic
transition compatibility and target duration. A structured dynamic program
efficiently identifies high-utility compositional priors
(\S\ref{subsec:hag}).

\item We develop an adaptive \emph{Skip Gater} that selects the amount of reuse
for each request using prompt--prior context and serving conditions, enabling context-dependent decisions
(\S\ref{subsec:skipper}).
\end{denseitemize}

Our evaluations across
AudioLDM2 and Stable Audio Open show that \sysname
achieves $1.48$--$3.74\times$ generation speedup with only ${\sim}$1K cached
audios while preserving or improving perceptual quality, without
retraining the model (\S\ref{sec:eval}).

\section{Related Work}
\label{sec:background}

\textbf{Efficient audio diffusion.}
Faster samplers reduce sampling steps without modifying the model, via deterministic sampling~\citep{song2021ddim} or higher-order ODE solvers~\citep{lu2022dpmsolverpp}. Distillation trains a student to match the
teacher in fewer steps, by progressively halving the step count~\citep{salimans2022progressive} or learning a consistency mapping to the trajectory endpoint~\citep{pmlr-v202-song23a}. Caching methods reuse intermediate features or model outputs across adjacent timesteps to skip computation~\citep{ma2023deepcache, liu2025timestep}. Audio-specific methods include consistency distillation for one- or few-step T2A
generation~\citep{bai2024consistencytta, liu2024audiolcm, saito2025soundctm},
step and layer distillation for music generation~\citep{novack2025presto},
adversarial post-training~\citep{novack2025arc}, and rectified- and mean-flow
T2A models~\citep{liu2025flashaudio, li2025meanaudiofastfaithfultexttoaudio}. However, these methods
largely adapt designs originally developed for image generation, rather than
exploiting properties specific to audio. \sysname is instead an audio-centric
system, built around the temporally localized and recurring structure of sound.

\textbf{RAG for diffusion models.}
RAG has been explored in T2A primarily to improve generation quality and generalization. Re-AudioLDM~\citep{yuan2024} retrieves relevant text--audio pairs to condition generation of rare and unseen sound events, while Audiobox TTA RAG~\citep{yang25h_interspeech} conditions on retrieved audio for zero- and few-shot generation. 
Recent work, including
NIRVANA~\citep{agarwal2024nirvana} and MoDM~\citep{xia2025modm}, shows that generation requests can be accelerated by warm-starting from a similar previous request, motivated by core image editing algorithms ~\cite{meng2022sdedit}. These methods, however, reuse a single retrieved sample for image generation. \sysname instead extends to multi-sample compositional reuse for T2A generation.

\section{\sysname Design}
\label{sec:design}

\subsection{Motivation and Overview} 
\label{subsec:motivation}

\begin{figure}[t]
  \centering
  \begin{minipage}[b]{0.49\linewidth}
    \centering
    \includegraphics[width=.9\linewidth]{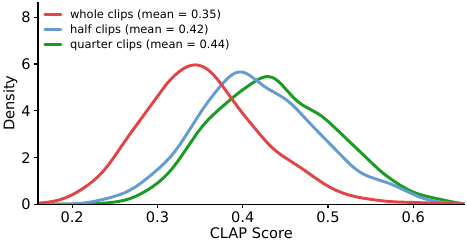}
    \caption{Whole and segmented audio often have many semantically similar neighbors.}
    \label{fig:motivation}
  \end{minipage}
  \hfill
  \begin{minipage}[b]{0.49\linewidth}
    \centering
    \includegraphics[width=.9\linewidth]{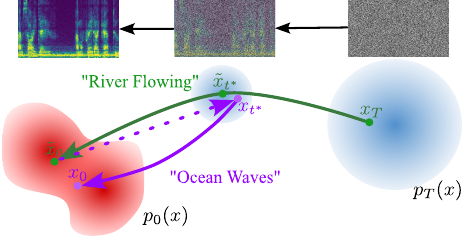}
    \caption{An intuitive geometric illustration of SoundWeaver's warm-starting.}
    \label{fig:math}
  \end{minipage}
  \label{fig:motivation-architecture-combined}
\vspace{-.4cm}
\end{figure}

Cross-example reuse is promising only if prior audio contains structure relevant
to new requests. In practice, such similarity is prevalent: as shown in
Fig.~\ref{fig:motivation}, many prompts have semantically aligned examples even
in a modest (1K) audio cache. However, the useful structure is often
\emph{temporally localized}: audios contain multiple events whose relevance varies over time,
so whole-clip retrieval can hide highly relevant intervals. Indeed, Fig.~\ref{fig:motivation} shows that progressively partitioning audios
from full clips into half- and quarter-length segments exposes substantially
stronger prompt--audio alignment. Furthermore, composing new audio from highly relevant segments of existing audios to form new audio naturally has a long history in traditional signal processing~\citep{hunt1996unit, Schwarz2007CorpusBasedCS}. However, the concatenated sequence is itself the final output, so poorly matched joins remain audible. This makes diffusion a natural complement as SoundWeaver uses audio composition only as a prior from which the diffusion model regenerates the final output.

\textbf{Theoretical Intuition for Warm-Start Acceleration.}
T2A diffusion models learn a reverse process $p_\theta(x_{t-1}\mid x_t,c)$
that transforms $x_T\sim\mathcal{N}(0,I)$ into a clean audio latent $x_0$
over $T$ denoising steps, conditioned on prompt $c$. The standard forward
process~\citep{song2021ddim} is
$
q(x_t\mid x_0)=\mathcal{N}\big(\sqrt{\bar\alpha_t}\,x_0,\,(1-\bar\alpha_t)I\big)$,
where $\bar\alpha_t$ decreases with $t$. Applying it to data
$x_0\sim p(x_0\mid c)$ gives the noised marginal $p_t(x_t\mid c)$.
Warm-starting encodes a reference $\tilde{x}_0$ into the model's latent space,
samples $x_{t^*}\sim q(x_{t^*}\mid\tilde{x}_0)$, and runs only the reverse
steps from $t^*$ to $0$, skipping $T-t^*$ steps. Let
$\Delta(\tilde{x}_0)=\mathbb{E}_{x_0\sim p(x_0\mid c)}\|\tilde{x}_0-x_0\|_2^2$
denote the expected distance between the reference and the target
distribution.

\begin{proposition}\label{prop:warmstart}
If the reverse process maps $p_t(x_t\mid c)$ to $p(x_0\mid c)$, its output
distribution $\hat{p}(x_0\mid c)$ when initialized from
$q(x_t\mid\tilde{x}_0)$ satisfies
\begin{equation}
\small
\begin{aligned}
D_{\mathrm{KL}}\big(\hat{p}(x_0\mid c)\,\|\,p(x_0\mid c)\big)
&\le D_{\mathrm{KL}}\big(q(x_t\mid\tilde{x}_0)\,\|\,p_t(x_t\mid c)\big) \le \frac{\bar\alpha_t}{2(1-\bar\alpha_t)}\,\Delta(\tilde{x}_0).
\end{aligned}
\end{equation}
\end{proposition}
The first inequality follows from the data processing inequality and the
second from the convexity of KL divergence
(proof in Appendix~\ref{app:proof}). Proposition~\ref{prop:warmstart} yields
three implications. First, the bound decays with the signal-to-noise ratio
$\mathrm{SNR}_t=\bar\alpha_t/(1-\bar\alpha_t)$: noising more deeply tolerates
a worse reference, but skips fewer steps. Second, for an error tolerance
$\epsilon$, the shallowest admissible warm-start satisfies
$\mathrm{SNR}_{t^*}\le 2\epsilon/\Delta(\tilde{x}_0)$, so a better reference
permits skipping more steps, motivating improving the prior itself (\S\ref{subsec:hag}).
Third, $\Delta(\tilde{x}_0)$ is unobservable, since it depends on the unknown
$p(x_0\mid c)$; SoundWeaver therefore treats prompt-alignment scores as
proxies for reference quality and learns the mapping from these signals to
$t^*$ (\S\ref{subsec:skipper}).


\begin{figure}[t]
  \centering
  \centerline{\includegraphics[width=\linewidth]{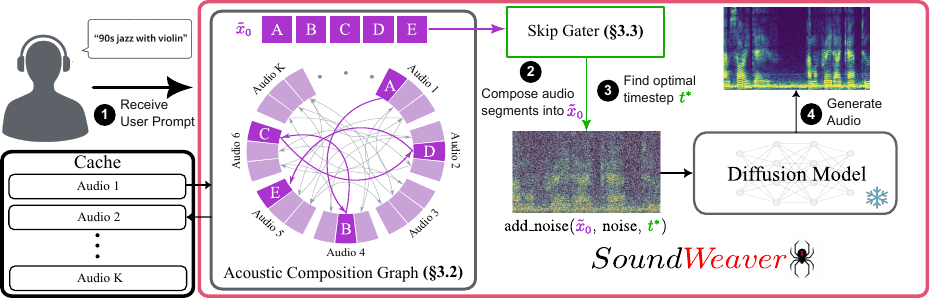}}
  \vspace{-.1cm}
  \caption{\sysname design overview and request execution flow.}
  \label{fig:architecture}
  \vspace{-.4cm}
\end{figure}

\textbf{Design Challenges.} 
Guided by this observation, we aim to improve the acoustic prior
\emph{before} deciding how aggressively to warm-start from it (Fig.~\ref{fig:architecture}).  
However, this formulation raises two key questions: \emph{(i)} how do we construct a
useful acoustic prior $\hat{x}_0$, especially when no single previous cache
matches the request in duration or content, and \emph{(ii)} how much of that prior can be
preserved for a given request? \sysname addresses these questions with an
\emph{Acoustic Composition Graph} (ACG) that composes reusable acoustic
structure across cached audios (\S\ref{subsec:hag}), and an adaptive
\emph{Skip Gater} that selects the warm-start timestep $t^*$ (\S\ref{subsec:skipper}).

\subsection{Acoustic Composition Graph}
\label{subsec:hag}

To warm-start T2A generation, \sysname must construct a semantically relevant
acoustic prior $\hat{x}_0$ that reflects the target prompt over the
requested duration. Retrieving the single most semantically similar cache assumes the same audio provides useful acoustic structure throughout the generation. In practice, however, the usefulness of a cached
audio can vary substantially over time: an audio may contain short intervals
that are highly relevant to the target prompt alongside irrelevant content, and
its duration may not match the requested duration. Instead, different parts of a prompt may be
better represented by different temporal regions of different caches.
Thus, useful acoustic structure need not align with the boundaries of any
single cached example.

Hence, \sysname constructs $\hat{x}_0$ compositionally from segments across
cached audios. Though, this introduces a second requirement: these segments originating from different audios must be joined together
without arbitrary acoustic discontinuities. We capture both requirements with an
\emph{Acoustic Composition Graph} (ACG), whose nodes correspond to cached audio
segments and whose edges represent possible transitions at a desired level of
acoustic similarity. Given a prompt and target duration, constructing the prior becomes a structured path-search problem over this graph where the path length is constrained by the requested duration.

\textbf{Compositional Acoustic Graph Construction.}
We divide each cached audio into segments of $L$ consecutive audio frames,
which form the nodes of the ACG. To determine which segments can be connected,
we need a representation that captures acoustic similarity at their boundaries.
Direct waveform comparison is poorly suited for identifying alternative
continuations since perceptually similar sounds can differ substantially in
sample-level phase, noise, and amplitude. Semantic embeddings are also
insufficient: they capture how two sounds convey similar
content, but not whether they occupy similar local acoustic states. We, therefore, take a cheap tokenizer approach and represent a frame, $f$, at the start of each segment using EnCodec ~\citep{defossez2022highfi}. EnCodec encodes $f$ using
residual vector quantization (RVQ) with $K$ codebooks: the first codebook
quantizes the frame's latent representation, while each subsequent codebook
quantizes the residual left by the preceding ones. A frame is, thus, represented
by a token sequence $(c_1,\ldots,c_K)$ that progressively refines its acoustic
representation. We denote by
$\mathrm{pre}_d(f)=(c_1,\ldots,c_d)$ the prefix formed by the first $d$
codebooks, which provides a natural hierarchy of acoustic similarity from
coarse to fine.

From a segment $u$, a path can continue to the segment that follows $u$ in its
source audio, which we denote $u^{+}$. It can also switch to another segment
$v$ whose first frame has the same $d$-token prefix as the first frame of
$u^{+}$.
Specifically, writing $f_s$ for the first frame of segment $s$, we admit an edge
\begin{equation}
  u \to v \iff \mathrm{pre}_d(f_v) = \mathrm{pre}_d(f_{u^{+}}).
  \label{eq:edge}
\end{equation}
Matching these boundary-frame prefixes makes $v$ a candidate which is an acoustically
compatible continuation of $u$. Smaller $d$ admits more cross-audio
transitions with coarser acoustic matching, while larger $d$ admits fewer
transitions with stricter acoustic similarity. To identify these matches
efficiently without comparing every pair of segments, \sysname stores the
prefix of each segment's first frame in a prefix tree.

\textbf{Relevance- and Continuity-Aware Path Search.}
The graph specifies which segments can be connected. We next determine which path best matches the new prompt. We propose an approach inspired by traditional audio processing ~\citep{hunt1996unit, Schwarz2007CorpusBasedCS} that generates utterances by searching audio segments that minimize a target cost (match to the desired output) plus a concatenation cost (penalizing perceptible boundary jumps) to produce highly natural audio. 

We employ FLAM~\citep{pmlr-v267-wu25ab}, a dual text--audio semantic encoder that specializes in scoring short sliding windows against the prompt text. Let $\rho(v)$ denote the
aggregate prompt-alignment score of segment $v$. To prevent audible boundary artifacts, \sysname
precomputes an acoustic distance penalty $\omega(u,v)$ for each candidate edge $u \to v$ offline, using the codec's continuous latent embedding $z_f$:
\begin{equation}
  \omega(u, v) =
  -\frac{1}{\tau}
  \bigl\lVert
  z_{\mathrm{last}(u)}
  -
  z_{\mathrm{last}(\mathrm{pred}(v))}
  \bigr\rVert^2,
  \label{eq:transition}
\end{equation}
where $\mathrm{pred}(v)$ denotes the segment immediately preceding $v$ in its
source audio, and $\tau$ is the median distance between adjacent
frames. The score does not prohibit cross-audio transitions; rather, it
downweights substitutions for which the acoustic context preceding $v$ differs
substantially from the end of $u$. Because audio embeddings are materialized
offline and cross-audio penalties can be precomputed offline, online
scoring requires only a single prompt text embedding and one matrix-vector dot
product across candidate segments.

For a new generation request with target duration $T$, \sysname searches for a
path of $n = \lceil rT / L \rceil$ segments, where $r$ is the codec frame rate,
selecting the length-$n$ path that jointly maximizes semantic relevance and
acoustic continuity:
\begin{equation}
  \max_{v_1 \to \cdots \to v_n}
  (1 - \alpha) \sum_{m=1}^{n} \rho(v_m)
  +
  \alpha \sum_{m=2}^{n} \omega(v_{m-1}, v_m),
  \label{eq:objective}
\end{equation}
where $\alpha \in [0,1]$ balances semantic relevance against smoothness. Since~\eqref{eq:objective} decomposes into segment-local rewards and first-order
transition scores, its global optimum can be found with dynamic programming
(DP):
\begin{equation}
  R_m(v) =
  (1 - \alpha) \rho(v)
  +
  \max_{u \to v}
  \bigl[
    R_{m-1}(u) + \alpha\,\omega(u,v)
  \bigr],
  \qquad
  R_1(v) = (1 - \alpha)\rho(v),
  \label{eq:dp}
\end{equation}
where $R_m(v)$ records the maximum cumulative score of an $m$-segment path
ending at $v$. Our DP solver scales linearly with the requested path length
$n$ and the number of admissible graph edges, requiring approximately
2\,ms per request in our evaluations (\S\ref{eval:ablation}).

\textbf{Global Reranking and Prior Assembly.}
The dynamic program optimizes an additive objective over local segment
relevance and pairwise transitions, but this objective does not directly
measure the semantic correspondence of the assembled clip as a whole. To
address this, we retain the top $M$ paths from the structured search
($M=16$ by default), assemble each candidate from its source waveform
segments, and rerank the resulting clips using whole-clip CLAP
similarity. This provides a global semantic signal
while evaluating only a small set of high-scoring candidates.

The top-ranked assembled candidate is the acoustic prior $\hat{x}_0$
passed to the Skip Gater (\S\ref{subsec:skipper}). Importantly,
codec tokens are used for indexing and structured search only. The prior is
assembled from the original cached waveforms, avoiding storage overhead and codec reconstruction
loss.

\subsection{Adaptive Warm-Start Gating}
\label{subsec:skipper}

Even for a semantically relevant composed prior, the appropriate amount of reuse varies across requests. Starting from a less-noised prior preserves more acoustic structure and enables skipping more denoising steps, but leaves less opportunity for the model to refine content that deviates from the target prompt.
Consequently, the appropriate skip depth depends on the prompt, composed prior, and model, and a fixed similarity threshold cannot capture this variation. Moreover, training a supervised predictor is often impractical since generation quality lacks reliable golden labels and request distributions evolve over time. We introduce a \emph{Skip Gater} that adaptively selects how many denoising steps to skip for each request.

\textbf{Contextual skip selection.}
We formulate warm-start selection as a contextual multi-armed bandit. Each arm
$k\in\mathcal{A}$ corresponds to a candidate skip depth, with arms spaced
uniformly in log-SNR. This choice follows Proposition~\ref{prop:warmstart}:
warm-start mismatch varies with SNR rather than raw diffusion timestep, so
log-SNR spacing provides approximately uniform multiplicative resolution over
the mismatch term. Given context $x$, containing embeddings of the prompt and
composed prior together with the denoising budget and serving state, the gater
selects the number of denoising steps to skip. We define the reward as:
\begin{equation}
r(x,k)
=
\Delta E(x,k)
+
\beta\widetilde{Q}(x,k),
\label{eq:gater_reward}
\end{equation}
where $\Delta E(x,k)\in[0,1]$ measures the serving benefit of skipping, such as
reduction in latency-SLO violation, and $\widetilde{Q}(x, k)$ measures
generation quality (e.g., CLAP score). The coefficient $\beta$ controls the quality--efficiency
trade-off (e.g., the developer's preference). Because the serving benefit depends on the current load, the same
prior can receive a conservative skip under light load and a more aggressive
one when latency reduction is valuable.

Adjacent skip actions are strongly related, so we should avoid learning an independent
model for each arm. To this end, we parameterize the expected reward with a
Bayesian linear model over the tensor product of a context representation
$\phi(x)$ and a continuous arm basis $\psi(k)$:
\begin{equation}
\widehat{R}_w(x,k)
=
\left\langle
\phi(x)\otimes\psi(k),w
\right\rangle,
\qquad
w\sim\mathcal{N}(m,\Sigma).
\label{eq:bandit_model}
\end{equation}
The arm basis is defined over log-SNR, allowing observations from one skip
level to inform nearby actions. We use Thompson sampling from the Gaussian
posterior during exploration and the posterior mean for deterministic
user-facing inference.

\textbf{Learning from noisy and heterogeneous feedback.}
Absolute quality scores $Q(x, k)$ are difficult to calibrate across prompts: users rarely provide reliable scalar ratings, and automatic proxies such as CLAP provide only
an imperfect proxy for perceptual quality. Fortunately, production systems like ChatGPT already collect preference-style feedback, such as
 thumbs up/down ratings or pairwise comparisons~\citep{Yu_2025}. Thus we
convert the observed quality of action into within-prompt percentile ranks
$\widetilde{Q}(x,k)\in[0,1]$. 
But still, skip sensitivity varies substantially across prompts. Our analysis of 1k randomly sampled AudioCaps prompts shows that CLAP variance spans nearly two orders of magnitude: simple prompts (e.g., \emph{``Water runs continuously''}) retain quality across wide skip ranges, whereas semantically rich prompts (e.g., \emph{``Dog barking while a cat is meowing...''}) exhibit rapid degradation. To focus learning on decisions that matter, we weight parameter updates by the prompt's quality variance:
\begin{equation}
    w_p = \frac{\sigma_p^2}{\bar{\sigma}^2 + \epsilon},
\end{equation}
where $\sigma_p^2$ is the quality variance across observed skip actions for prompt $p$, and $\bar{\sigma}^2$ is the mean variance across all prompts. This emphasizes requests for which
skip depth materially affects generation quality. The posterior is initialized
from offline action sweeps and updated asynchronously from logged outcomes, so learning does not add latency to the generation path. Specifically, when a generation receives negative feedback, we log the prompt and chosen skip depth, then asynchronously regenerate the prompt at every arm off the serving path. This yields a full within-prompt quality profile from which we compute $\widetilde{Q}$ and $\sigma_p^2$.



\begin{table*}[t]
\centering
\caption{
Objective and subjective generation-quality metrics and inference latency
across different audio-generation backbones, baselines, and ablations.
$^\emph{S}$~synthetic-audio cache;
$^\emph{R}$~real-audio cache.
}
\label{tab:quality}
\vspace{4pt}

\small
\setlength{\tabcolsep}{3.5pt}
\renewcommand{\arraystretch}{1.08}

\begin{adjustbox}{max width=\textwidth}
\begin{tabular}{l l c c c c c c c c c}
\toprule

\multicolumn{1}{c}{\multirow{2}{*}[-0.6ex]{\textbf{Backbone}}}
& \multicolumn{1}{c}{\multirow{2}{*}[-0.6ex]{\textbf{Method}}}
& \multicolumn{4}{c}{\textbf{Objective}}
& \multicolumn{3}{c}{\textbf{Subjective}}
& \multicolumn{2}{c}{\textbf{Latency}} \\

\cmidrule(lr){3-6}
\cmidrule(lr){7-9}
\cmidrule(lr){10-11}

&
& \textbf{FD} $\downarrow$
& \textbf{KL} $\downarrow$
& \textbf{CLAP} $\uparrow$
& \textbf{CCDS} $\downarrow$
& \textbf{PS} $\uparrow$
& \textbf{OVL} $\uparrow$
& \textbf{REL} $\uparrow$
& \textbf{Value (s)} $\downarrow$
& \textbf{Speedup} $\uparrow$ \\

\midrule

\multirow[c]{7}{*}{
  \shortstack{\textbf{AudioLDM 2}}
}
& Vanilla
& 30.00
& 1.56
& 0.41
& 0.72
& --
& --
& --
& 3.95
& -- \\

& TeaCache
& 43.86
& 1.87
& 0.36
& 0.74
& 0.33
& --
& --
& 2.67
& $1.5\times$ \\

& DeepCache
& 39.68
& 1.73
& 0.38
& 0.78
& 0.30
& --
& --
& 1.68
& $2.3 \times$ \\

& MoDM
& 25.54
& 1.78
& 0.39
& 0.78
& 0.37
& --
& --
& 2.81
& $1.4\times$ \\

& \cellcolor{oursbg}\sysname~$^\emph{S}$
& \cellcolor{oursbg}23.81
& \cellcolor{oursbg}1.41
& \cellcolor{oursbg}0.44
& \cellcolor{oursbg}0.75
& \cellcolor{oursbg}0.41
& \cellcolor{oursbg}0.43
& \cellcolor{oursbg}0.61
& \cellcolor{oursbg}2.12
& \cellcolor{oursbg}$1.9\times$ \\

& \cellcolor{oursbg}\sysname~$^\emph{R}$
& \cellcolor{oursbg}20.53
& \cellcolor{oursbg}1.50
& \cellcolor{oursbg}0.44
& \cellcolor{oursbg}0.72
& \cellcolor{oursbg}0.40
& \cellcolor{oursbg}0.68
& \cellcolor{oursbg}0.70
& \cellcolor{oursbg}2.67
& \cellcolor{oursbg}$1.5\times$ \\

& \quad w/o ACG
& 22.63
& 1.68
& 0.41
& 0.82
& 0.37
& --
& --
& 2.62
& $1.5\times$ \\

& \quad w/o SG
& 24.08
& 1.45
& 0.42
& 0.79
& 0.36
& --
& --
& 2.63
& $1.5\times$ \\

\midrule

\multirow[c]{7}{*}{
  \shortstack{\textbf{Stable Audio}\\
  \textbf{Open}}
}
& Vanilla
& 35.04
& 1.98
& 0.29
& 0.65
& --
& --
& --
& 14.66
& -- \\

& TeaCache
& 41.38
& 2.00
& 0.27
& 0.68
& 0.43
& --
& --
& 8.45
& $1.7\times$ \\

& SmoothCache
& 36.27
& 2.01
& 0.27
& 0.67
& 0.42
& --
& --
& 12.61
& $1.2 \times$ \\

& MoDM
& 36.99
& 1.88
& 0.29
& 0.67
& 0.37
& --
& --
& 8.41
& $1.7\times$ \\

& \cellcolor{oursbg}\sysname~$^\emph{S}$
& \cellcolor{oursbg}27.88
& \cellcolor{oursbg}1.72
& \cellcolor{oursbg}0.36
& \cellcolor{oursbg}0.69
& \cellcolor{oursbg}0.50
& \cellcolor{oursbg}0.30
& \cellcolor{oursbg}0.43
& \cellcolor{oursbg}6.10
& \cellcolor{oursbg}$2.4\times$ \\

& \cellcolor{oursbg}\sysname~$^\emph{R}$
& \cellcolor{oursbg}27.48
& \cellcolor{oursbg}1.72
& \cellcolor{oursbg}0.34
& \cellcolor{oursbg}0.64
& \cellcolor{oursbg}0.42
& \cellcolor{oursbg}0.70
& \cellcolor{oursbg}0.59
& \cellcolor{oursbg}3.92
& \cellcolor{oursbg}$3.7\times$ \\

& \quad w/o ACG
& 29.30
& 1.83
& 0.33
& 0.81
& 0.38
& --
& --
& 4.39
& $3.3\times$ \\

& \quad w/o SG
& 27.96
& 1.65
& 0.33
& 0.72
& 0.43
& --
& --
& 4.69
& $3.1\times$ \\

\bottomrule
\end{tabular}
\end{adjustbox}
\end{table*}
\section{EVALUATION}
\label{sec:eval}

\subsection{Experimental Setup}

We evaluate T2A diffusion backbones across architectures and
samplers: AudioLDM2 (1.1B)~\citep{audioldm2-2024taslp} with
DDIM~\citep{song2021ddim}, and Stable Audio Open
(1.3B)~\citep{evans2025stableaudioopen} with
DPM-Solver++~\citep{lu2022dpmsolverpp}. We use
50 sampling steps by default, and perform studies on different sampling steps (\S\ref{eval:ablation}). All experiments were done on an NVIDIA A100 GPU. Following prior work~\citep{liu2023audioldm,audioldm2-2024taslp},
we evaluate on the AudioCaps test set, randomly selecting one caption per clip
as the generation prompt.

We consider two cache sources: (i) \emph{real-audio cache} taken from open-access Clotho~v2 audios ~\cite{drossos2019clothoaudiocaptioningdataset}, and (ii) the \emph{synthetic cache} containing audio generated by the evaluated model from the corresponding Clotho captions. Thus, the two audio sources differ only in quality. All ablation studies and analysis use the \emph{real-audio cache} variant unless stated otherwise. Additional implementation details such as image to audio porting procedures are provided in
Appendix~\ref{app:hyperparameters}.

\textbf{Baselines.}
We compare \sysname against the following approaches:
\begin{denseitemize}

    \item \emph{Vanilla}: Standard model inference from gaussian noise

    \item \emph{DeepCache}~\citep{ma2023deepcache}: Accelerates diffusion U-Nets by reusing intermediate layer outputs at fixed step intervals. We evaluate it on our U-Net backbone, AudioLDM2.

    \item \emph{SmoothCache}~\citep{liu2025smoothcache}: Accelerates Diffusion Transformers (DiTs) by reusing intermediate layer outputs under calibrated error bounds. We evaluate it on our DiT backbone, Stable Audio Open.

    \item \emph{TeaCache}~\citep{liu2025timestep}: Accelerates diffusion models by reusing final model outputs when input differences indicate negligible change.

    \item \emph{MoDM}~\citep{xia2025modm}:Accelerates diffusion models by retrieving a semantically similar reference to warm-start diffusion from an intermediate step.

\end{denseitemize}
\textbf{Metrics.}
We evaluate generation quality using human judgments, LLMs, and objective metrics:
\begin{denseitemize}

\item \emph{Human evaluation}: Winrate as a fraction against Vanilla in randomized pairwise comparisons for \emph{Overall Quality (OVL)} and prompt \emph{Relevance (REL)} following prior work ~\cite{liu2023audioldm, audioldm2-2024taslp}. We report mean values from 22 random participants across 12 audio-prompt pairs each.

\item \emph{LLM-based evaluation}: Winrate as a fraction against Vanilla in randomized pairwise comparisons for a \emph{Preference Score (PS)} following existing advances ~\citep{wang2025speechllmasjudgesgeneralinterpretablespeech, zheng2023judging}. We report mean values across 100 audio pairs. For scalable evaluation, we use Gemini Flash 3.7 as a pairwise audio judge. Evaluation prompt is provided in the AI Use Statement.

\item \emph{Objective metrics}: \emph{CLAP}~\citep{CLAP2023} measures text--audio semantic alignment; \emph{KL divergence (KL)}~\citep{kl_div} compares acoustic-event posteriors; \emph{Fr\'echet Distance (FD)}~\citep{frechet1906sur}, measures distributional similarity to reference audio; and \emph{CLAP Conditional Diversity Score (CCDS)} ~\cite{novack2025arc} measures pairwise similarity of the outputs generated using two different seeds. 

\end{denseitemize}

\subsection{Main Results}
\label{eval:e2e}

\begin{figure*}[t]
  \centering

  \begin{minipage}[t]{0.49\textwidth}
    \centering
    \includegraphics[width=\linewidth]{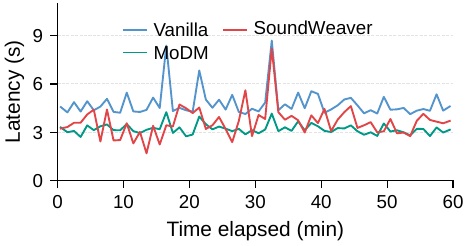}
    \centerline{(a) AudioLDM 2}\medskip
  \end{minipage}
  \hfill
  \begin{minipage}[t]{0.49\textwidth}
    \centering
    \includegraphics[width=\linewidth]{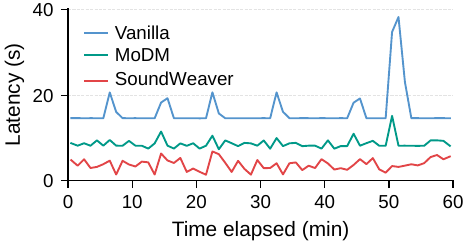}
    \centerline{(b) Stable Audio Open}\medskip
  \end{minipage}
\vspace{-.3cm}
  \caption{\sysname improves online serving latency in our one-hour deployments.}
  \label{fig:online-lat}
  \vspace{-.5cm}
\end{figure*}

\textbf{\sysname improves latency and quality.}
As shown in Table~\ref{tab:quality}, \sysname achieves 1.48--3.74$\times$ latency speedups on top of each model's native sampler while consistently improving quality. The real-audio cache yields the largest quality gains, which we attribute to the richer perceptual detail in real audios, and
comparable CCDS indicates that output diversity is preserved. These gains
hold under online serving (Fig.~\ref{fig:online-lat}): on a one-hour deployment driven by request traces
from DiffusionDB~\citep{diffusion-db} with audio prompts and durations used from WavCaps ~\cite{mei2024wavcaps}, and arrival rates scaled to our
hardware, warm-starting reduces per-request NFEs and yields 1.06--11.82$\times$ end-to-end speedups.

\begin{figure}[t]
  \centering
\begin{minipage}[t]{0.32\linewidth}
    \centering
    \includegraphics[width=\linewidth]{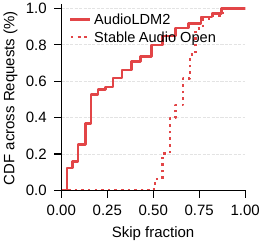}
    \vspace{-.8cm}
    \caption{\sysname adaptively skips denoising steps.}
    \label{fig:cdf_models}
  \end{minipage}
  \hfill
  \begin{minipage}[t]{0.65\linewidth}
    \centering
    \includegraphics[width=\linewidth]{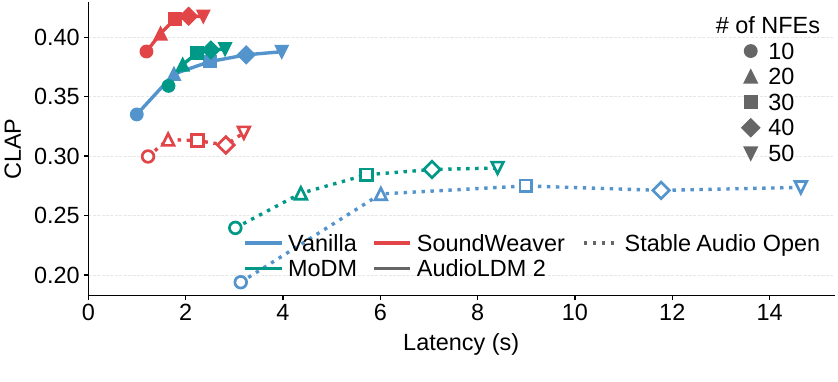}
    \vspace{-.8cm}
    \caption{SoundWeaver unlocks new quality-efficiency frontiers across sampling budgets (NFEs) in a training-free manner.}
    \label{fig:latency_models}
  \end{minipage}
\vspace{-.2cm}
\end{figure}

\textbf{\sysname adapts skip depth across requests.} Figure~\ref{fig:cdf_models} shows substantial variation in the selected skip
ratios across prompts. This wide distribution confirms that the Skip Gater does not
collapse to a fixed policy, but adjusts the amount of reuse to each request;
the corresponding quality results in Table~\ref{tab:quality} show that these
latency savings are achieved without sacrificing generation quality. Stable Audio Open exhibits
higher skip ratios overall, largely because its sampler allocates more steps to
high noise regions. So, a warm start at the same SNR removes a larger fraction
of the trajectory than under AudioLDM2's DDIM schedule, consistent with
Proposition~\ref{prop:warmstart}.

\begin{table}[t]
  \centering

  \begin{minipage}[t]{0.49\linewidth}
    \centering
    \small
    \setlength{\tabcolsep}{5pt}
    \begin{tabular}{c c c}
      \toprule
      \textbf{Cache Size (Memory)}
      & \textbf{CLAP} $\uparrow$
      & \textbf{Latency (s)} $\downarrow$ \\
      \midrule
      100 (0.04 GB) & 0.38 & 3.47 \\
      500 (0.20 GB) & 0.39 & 3.05 \\
      1K  (0.39 GB) & 0.42 & 2.55 \\
      5K  (1.97 GB) & 0.43 & 2.04 \\
      \bottomrule
    \end{tabular}

    \centerline{\small (a) AudioLDM 2}
  \end{minipage}
  \hfill
  \begin{minipage}[t]{0.49\linewidth}
    \centering
    \small
    \setlength{\tabcolsep}{5pt}
    \begin{tabular}{c c c}
      \toprule
      \textbf{Cache Size (Memory)}
      & \textbf{CLAP} $\uparrow$
      & \textbf{Latency (s)} $\downarrow$ \\
      \midrule
      100 (0.04 GB) & 0.29 & 4.90 \\
      500 (0.20 GB) & 0.29 & 3.96 \\
      1K  (0.39 GB) & 0.30 & 3.49 \\
      5K  (1.97 GB) & 0.36 & 3.19 \\
      \bottomrule
    \end{tabular}

    \centerline{\small (b) Stable Audio Open}
  \end{minipage}

  \caption{Impact of cache pool size on SoundWeaver (WavCaps dataset \citep{mei2024wavcaps}).}
  \label{tab:cachesize}
  \vspace{-.2cm}
\end{table}

\begin{figure}[t]
  \centering

  \begin{minipage}[t]{0.65\linewidth}
    \centering

    \begin{minipage}[t]{0.49\linewidth}
      \centering
      \includegraphics[width=1.04\linewidth]{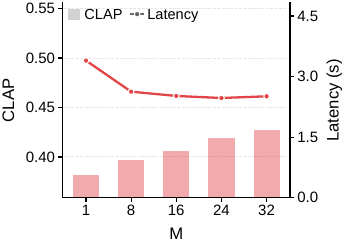}
      \centerline{\small (a) Impact of beam size.}
    \end{minipage}
    \hfill
    \begin{minipage}[t]{0.49\linewidth}
      \centering
      \includegraphics[width=1.04\linewidth]{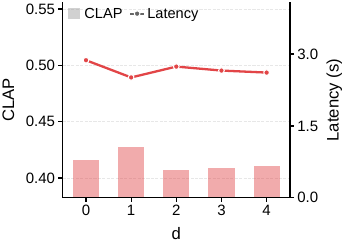}
      \centerline{\small (b) Impact of prefix depth.}
    \end{minipage}
    \vspace{-.1cm}
    \caption{Sensitivity analysis (AudioLDM 2).}
    \label{fig:sensitivity}
  \end{minipage}
  \hfill
  \begin{minipage}[t]{0.32\linewidth}
    \centering

    \includegraphics[width=1.04\linewidth]{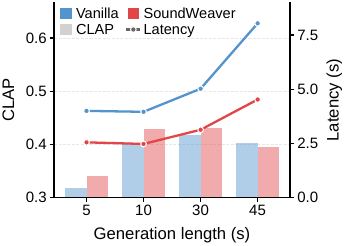}

    \caption{Impact of generation length (AudioLDM 2).}
    \label{fig:length-study}
  \end{minipage}
\vspace{-.6cm}
\end{figure}





\textbf{\sysname introduces negligible overhead.}
Prior construction and warm-start selection add only ${\sim}0.1$s per
request: ACG search takes a median of 68--79ms, while the Skip Gater requires
27--29ms. Together, this accounts for less than 3\% of vanilla generation
time on AudioLDM2 and less than 1\% on Stable Audio Open, while warm-starting
reduces overall latency by 32--73\% (Table~\ref{tab:quality}).

\subsection{Ablation Studies}
\label{eval:ablation}

\textbf{Performance Breakdown.}
We dive into the contributions of the Acoustic Composition Graph (ACG) and
adaptive Skip Gater (SG) in Table~\ref{tab:quality}. Removing ACG replaces
compositional trajectory search with whole-clip nearest-neighbor retrieval using CLAP score.
This consistently degrades generation quality; for example, FD increases from
20.53 to 22.63 on AudioLDM2 and from 27.48 to 29.30 on Stable Audio Open.
These results indicate that composing prompt-relevant segments across audios
produces stronger warm-start priors than relying on a single retrieved clip.
Removing SG instead replaces the contextual bandit with a fixed
similarity-threshold policy following prior warm-start approaches
~\citep{agarwal2024nirvana,xia2025modm}. At comparable serving latency, the
adaptive gater yields consistently better quality, showing the benefit of
request-dependent skip selection over a static policy.

\textbf{Impact of Sampling Budget (NFEs).}
Figure~\ref{fig:latency_models} evaluates \sysname across sampling budgets from 10 to 50
steps. \sysname remains effective even at low budgets, where relatively few
denoising steps are available to skip. The quality benefit is most pronounced
in this regime, suggesting that a structured acoustic prior is particularly
valuable when the sampler has limited opportunity to construct the target
audio from noise. As the sampling budget increases, the absolute opportunity
for skipping also grows, yielding larger latency savings.

\textbf{Impact of Cache Pool Size.}
Table~\ref{tab:cachesize} varies the cache from 100 to 5K audio clips.
Increasing cache size clearly improves both CLAP and latency. This indicates that larger caches provide more
useful acoustic structure for composition, though substantial benefits already
emerge at comparatively modest cache sizes. Meanwhile, \sysname does not require an exhaustive audio database to construct rich composite priors, ensuring low overhead.

\textbf{Impact of Candidate Size $M$.} Figure~\ref{fig:sensitivity}(a) varies $M\in\{1,8,16,24,32\}$, the number of
highest-scoring trajectories retained by graph search for global CLAP
reranking. Increasing $M$ improves generation quality by giving the reranker a
broader set of compositions from which to select, with gains gradually
saturating around $M=32$. We also observe lower end-to-end latency with larger
$M$, as the resulting higher-quality priors allow the Skip Gater to select more
aggressive warm starts. 

\textbf{Impact of Prefix Matching Depth $d$.} Figure~\ref{fig:sensitivity}(b) evaluates the EnCodec prefix depth $d \in \{0, 1, 2, 3, 4\}$, used to establish transition edges between segment boundaries (\S\ref{subsec:hag}). Setting $d=1$ where at least one token has to be matched achieves the best overall generation quality. Stricter prefix constraints ($d \ge 1$) overly restrict candidate branching, yielding a sparser graph that limits compositional reuse without improving transition naturalness. Whilst at $d=0$ creating edges without an acoustic filter is also harmful. End-to-end latency remains essentially flat across different values of $d$, as the dynamic programming search overhead is only $\sim$2\,ms and is completely dominated by downstream diffusion.

\textbf{Impact of Generation Length.}
Figure~\ref{fig:length-study} shows the impact of target audio duration across 5, 10, 30, and 45 seconds. On AudioLDM~2, \sysname maintains consistent CLAP improvements of ${\sim}0.01 \pm 0.02$ over vanilla generation across varied horizons, while providing substantial latency savings which increase with duration. For Stable Audio Open, latency and quality remains largely constant across durations due to its architecture generating fixed 45-second contexts natively before temporal truncation.



\section{CONCLUSION}
This paper introduces \sysname, a training-free framework for accelerating T2A diffusion
through cross-example acoustic structure reuse. For a new request, \sysname constructs a prompt-conditioned acoustic
prior from reusable segments of previously observed audio and warm-starts
diffusion from an intermediate noise level. Its acoustic graph enables
compositional reuse across audios while accounting for semantic relevance
and transition compatibility, and an adaptive Skip Gater determines how
aggressively the resulting prior can be reused for each request. SoundWeaver achieves a $1.48$--$3.74\times$ generation speedup while preserving or improving generation quality.

\subsection*{AI use statement}
\label{ai}
In this work, we used Gemini Flash 3.7 to get the preference score (PS) documented in Table~\ref{tab:quality}. Claude and OpenAI models were additionally used to assist in proposing and refining hypotheses; coding, designing, and critiquing methodology and experiments; formulating mathematical claims; and helping write the paper and proofs. The authors verified all AI-assisted work. We take responsibility for the final paper.

The prompt used to get the preference scores is as follows:

\begin{tcolorbox}[
    enhanced,
    breakable,
    colback=gray!4,
    colframe=black!75,
    arc=3pt,
    boxrule=0.6pt,
    title={\small\textbf{Gemini Audio Evaluation Prompt}},
    fonttitle=\small\bfseries,
    coltitle=white,
    attach boxed title to top left={yshift=-2mm, xshift=6mm},
    boxed title style={
        colback=black!80,
        colframe=black!80,
        arc=2pt,
        boxrule=0pt,
    },
    top=4mm,
    left=4mm,
    right=4mm,
    bottom=3mm,
]

You are an expert audio quality evaluator.
You will be given two audio clips generated from the same text prompt and need to choose the audio file that sounds better.  Your evaluation should consider
\textbf{consistency}, \textbf{intelligibility}, \textbf{naturalness},
\textbf{absence of artifacts}, \textbf{timbral richness}, \textbf{fidelity},
and overall \textbf{quality}. Avoid any position biases and ensure that the order in which the audio were presented 
does not influence your decision. Do not allow the duration of the audio to influence 
your evaluation. Be as objective as possible.

\smallskip

You should start with your evaluation by comparing the two audios and provide a short
rationale about which one you think is better. After providing your rationale, you should output 
the final verdict by strictly following this seven-point Likert scale:

\smallskip
Compare the two audio clips and score the difference on a scale from -3 to 3:
\begin{center}
\begin{tabular}{cl}
    \textbf{Score} & \textbf{Meaning} \\
    \hline
    $+3$ & Audio 1 is \underline{much} better than Audio 2 \\
    $+2$ & Audio 1 is \underline{noticeably} better \\
    $+1$ & Audio 1 is \underline{slightly} better \\
    $\phantom{+}0$ & Both are equal / indistinguishable \\
    $-1$ & Audio 2 is \underline{slightly} better \\
    $-2$ & Audio 2 is \underline{noticeably} better \\
    $-3$ & Audio 2 is \underline{much} better than Audio 1 \\
\end{tabular}
\end{center}

\smallskip

Respond strictly in this format:

\texttt{[Rationale]:} \textit{short rationale, less than 200 words}\\
\texttt{[Score]:} \textit{integer from} $-3$ \textit{to} $+3$

\end{tcolorbox}




\subsection*{Ethics statement}

SoundWeaver warm-starts generation from previously observed audio, so traces of cached audios may persist in outputs. We mitigate this by fundamentally using fine acoustic units through our ACG, conservatively tuning our Skip Gater to avoid excessive skips, in addition to reporting output diversity (CCDS) throughout. Our real-audio cache uses only Creative Commons--licensed
Clotho~v2 audios, and SoundWeaver remains highly effective with a fully synthetic cache generated by the model itself (Table~\ref{tab:quality}) as well. We encourage researchers and developers interested in SoundWeaver to only use audio licensed for reuse. Listening-study participants were anonymous volunteers.

\subsection*{Reproducibility statement}

SoundWeaver is training-free and built entirely on publicly available
pretrained models: AudioLDM~2 and Stable Audio Open as diffusion backbones,
the 24\,kHz EnCodec model for acoustic tokenization, and FLAM and CLAP for
prompt--audio scoring. Section~\ref{sec:design} specifies each component of
the method, including the edge-admission rule for the Acoustic Composition
Graph (Eq.~2), the transition penalty (Eq.~3), the path-search objective and
its dynamic-programming solution (Eqs.~4--5), whole-clip reranking, and the
Skip Gater's reward, Bayesian parameterization, and variance-weighted
updates (Eqs.~6--8). Appendix~\ref{app:hyperparameters} lists all implementation details
and hyperparameters, including samplers, classifier-free guidance scales,
segment length $L$, prefix depth $d$, trade-off weight $\alpha$, candidate
count $M$, and how the MoDM and baselines were adapted to audio.
Appendix~\ref{app:proof} states the assumptions behind Proposition~1 and
gives its complete proof. All datasets are public: we evaluate on the
AudioCaps test set, build real and synthetic caches from Clotho~v2, use
WavCaps for the cache-size study and serving prompts, and derive request
arrival traces from DiffusionDB; Section~\ref{sec:eval} describes prompt
sampling, cache construction, and all evaluation metrics. The full prompt
used for the LLM-based preference judge is included in the AI Use Statement.

\bibliography{iclr2027_conference}
\bibliographystyle{iclr2027_conference}

\newpage
\appendix

\section{Implementation Details and Hyperparameters}
\label{app:hyperparameters}

\textbf{Diffusion Model and Sampler Configurations.}
We configure \sysname across two primary text-to-audio diffusion backbones: AudioLDM~2 (1.1B)~\citep{audioldm2-2024taslp} and Stable Audio Open (1.3B)~\citep{evans2025stableaudioopen}. Both backbones default to 50 sampling steps ($N_{\mathrm{step}} = 50$). AudioLDM~2 uses the DDIM sampler \citep{song2021ddim} with a classifier-free guidance (CFG) scale of 3.5, while Stable Audio Open employs DPM-Solver++ \citep{lu2022dpmsolverpp} with a CFG scale of 7.0.

\textbf{Graph Construction and Path Search.}
To construct the Acoustic Composition Graph  (ACG), each cached audio is encoded using the pre-trained 24\,kHz, 12\,kbps EnCodec model 
\citep{defossez2022highfi} The tokenized sequence is split into non-overlapping segments of length $L = 150$ frames, each serving as an individual graph node. A directed edge is instantiated between segment $u$ and $v$ whenever the boundary frames share an identical EnCodec residual vector quantization prefix of depth $d = 1$. 

During structured dynamic programming search, the trade-off hyperparameter balancing prompt semantic relevance against the continuous latent transition penalty is set to $\alpha = 0.25$. The top $M = 32$ candidate paths extracted by dynamic programming are fully assembled from raw source waveforms and reranked using whole-clip CLAP similarity to determine the final acoustic prior $\tilde{x}_0$.

\textbf{Baseline Adaptations to Audio.}
\begin{denseitemize}
\item \textbf{DeepCache Adaptation:} The algorithm was directly implemented over our baselines and hyperparameters were re-tuned for the audios setting.

\item \textbf{SmoothCache Adaptation:} SmoothCache was applied using the authors' released implementation without modifying the caching algorithm. We adapted its cache and calibration hooks to Stable Audio Open's \texttt{StableAudioDiTBlock} and recalibrated the layer-wise caching schedule on 10 AudioCaps training captions disjoint from evaluation. We set the relative-L1 threshold to $\alpha=0.18$, resulting in 26\% of sub-layer invocations being cached. All sampling settings, prompts, and noise seeds were identical to the other baselines.

\item \textbf{TeaCache Adaptation:} TeaCache was ported from the authors' official TangoFlux implementation to AudioLDM2 and Stable Audio Open without modifying the caching algorithm. Because the released rescaling coefficients are FLUX-specific, we re-fit the degree-4 polynomial on 70 AudioCaps training captions disjoint from evaluation and re-tuned the threshold to $\delta=0.003$ for AudioLDM2 (30\% skipped steps) and $\delta=0.05$ for Stable Audio Open (43\% skipped steps), matching MoDM's mean latency. On Stable Audio Open, sampler noise makes the resulting policy effectively timestep-based rather than input-adaptive. All sampling settings and prompts were identical to the other baselines.

\item \textbf{MoDM Adaptation:} To adapt the vision-based MoDM framework to audio generation, we replace visual similarity backbones with a CLAP audio similarity backbone. We re-calibrated the static, step-wise skip policy using AudioCaps training prompts. To ensure fair comparison with SoundWeaver we utilized a single backbone model rather than the proposed mixture-of-diffusion-models for cache generation and utilization. 
\end{denseitemize}

\textbf{Subjective Evaluation}
\begin{denseitemize}
\item \emph{Preference Score}: we compare our generations with the vanilla generations of the models. To get the final win rate, we convert each positive score to be a win, and each negative score to be a loss, while ties are discarded and tally up the wins for each model.

\item \emph{Human Evaluation}: Twenty-two volunteers participated in the evaluation. Each survey included 12 randomly selected prompts from the AudioCaps test set. For each prompt, participants compared two audio samples: one generated by either \sysname$^{S}$ or \sysname$^{R}$ and one generated by the vanilla baseline. The two \sysname{} variants were equally represented, and audio presentation order was randomized. Participants answered two questions per prompt: ``Which audio has better perceptual quality (naturalness, clarity, and fewer audible artifacts)?'' for overall quality (OVL), and ``Which audio best matches the prompt?'' for prompt relevance (REL). We report OVL and REL as the fraction of pairwise comparisons won by each model.

\end{denseitemize}

\section{Proof of Proposition~\ref{prop:warmstart}}
\label{app:proof}

Fix a prompt $c$ and a timestep $t$. Let $K_t$ denote the Markov kernel
induced by running the model's reverse sampler from time $t$ to $0$, so
that initializing the sampler with a distribution $\mu$ over $x_t$
yields the output distribution $K_t\mu$ over $x_0$. This covers both
stochastic samplers and deterministic ODE samplers such as
DDIM~\citep{song2021ddim}, for which $K_t$ is the push forward under a
deterministic map. By assumption, the reverse process is exact on the
noised data marginal:
\begin{equation}
K_t\, p_t(\cdot\mid c) = p(\cdot\mid c).
\end{equation}
Warm-starting from reference $\tilde{x}_0$ initializes the sampler with
$q(x_t\mid\tilde{x}_0)$, so the output distribution is
$\hat{p}(\cdot\mid c) = K_t\, q(\cdot\mid\tilde{x}_0)$.

\textbf{First inequality.}
By the data processing inequality, applying the same Markov kernel to
two distributions cannot increase the KL divergence between them:
\begin{equation}
D_{\mathrm{KL}}\big(\hat{p}(x_0\mid c)\,\|\,p(x_0\mid c)\big)
= D_{\mathrm{KL}}\big(K_t\, q(\cdot\mid\tilde{x}_0)\,\|\,K_t\, p_t(\cdot\mid c)\big)
\le D_{\mathrm{KL}}\big(q(x_t\mid\tilde{x}_0)\,\|\,p_t(x_t\mid c)\big).
\end{equation}

\textbf{Second inequality.}
The noised marginal is a mixture of Gaussians,
\begin{equation}
p_t(x_t\mid c) = \int q(x_t\mid x_0)\, p(x_0\mid c)\, dx_0.
\end{equation}
Since KL divergence is convex in its second argument,
\begin{equation}
D_{\mathrm{KL}}\big(q(x_t\mid\tilde{x}_0)\,\|\,p_t(x_t\mid c)\big)
\le \mathbb{E}_{x_0\sim p(x_0\mid c)}
\Big[ D_{\mathrm{KL}}\big(q(x_t\mid\tilde{x}_0)\,\|\,q(x_t\mid x_0)\big) \Big].
\end{equation}
Both distributions inside the expectation are Gaussians with shared
covariance $(1-\bar\alpha_t)I$ and means $\sqrt{\bar\alpha_t}\,\tilde{x}_0$
and $\sqrt{\bar\alpha_t}\,x_0$. The KL divergence between Gaussians with
equal covariance $\sigma^2 I$ and means $m_1, m_2$ is
$\|m_1 - m_2\|_2^2 / (2\sigma^2)$, so
\begin{equation}
D_{\mathrm{KL}}\big(q(x_t\mid\tilde{x}_0)\,\|\,q(x_t\mid x_0)\big)
= \frac{\bar\alpha_t}{2(1-\bar\alpha_t)}\,\|\tilde{x}_0 - x_0\|_2^2.
\end{equation}
Taking the expectation over $x_0\sim p(x_0\mid c)$ gives
\begin{equation}
D_{\mathrm{KL}}\big(q(x_t\mid\tilde{x}_0)\,\|\,p_t(x_t\mid c)\big)
\le \frac{\bar\alpha_t}{2(1-\bar\alpha_t)}\,\Delta(\tilde{x}_0),
\end{equation}
which completes the proof. \qed

\textbf{Shallowest admissible warm-start.}
Requiring the bound to be at most a tolerance $\epsilon$ gives
$\mathrm{SNR}_t\,\Delta(\tilde{x}_0)/2 \le \epsilon$, i.e.,
$\mathrm{SNR}_t \le 2\epsilon/\Delta(\tilde{x}_0)$. Since
$\mathrm{SNR}_t$ decreases monotonically in $t$, the set of admissible
timesteps is an interval $[t_{\min}, T]$, and the shallowest admissible
warm-start $t^*$ is its smallest element. A smaller $\Delta(\tilde{x}_0)$
raises the SNR threshold and therefore lowers $t^*$, skipping more steps.

\textbf{Remark: the bound and output diversity.}
Expanding $\Delta$ around the conditional mean
$\mu_c = \mathbb{E}[x_0\mid c]$ gives
\begin{equation}
\Delta(\tilde{x}_0) = \|\tilde{x}_0 - \mu_c\|_2^2
+ \mathbb{E}_{x_0\sim p(x_0\mid c)}\|x_0 - \mu_c\|_2^2.
\end{equation}
The second term is the intrinsic variance of the target distribution and
does not depend on the reference. Hence the bound remains positive even
for an ideal reference: initializing from a single Gaussian
$q(x_t\mid\tilde{x}_0)$ rather than the full mixture $p_t(x_t\mid c)$
concentrates outputs around the reference. This quantifies the
diversity cost of warm-starting and motivates measuring output diversity
(CCDS) alongside quality in \S\ref{sec:eval}.

\end{document}